\documentstyle[12pt]{article}\newcommand{\nc}{\newcommand}
\nc{\be}{\beta}
\nc{\th}{\theta} \nc{\dl}{\delta} \nc{\ph}{\varphi}
\nc{\ep}{\varepsilon}
\nc{\ov}{\over} \nc{\ra}{\rightarrow} \nc{\hf}{{1\ov2}}
\nc{\bq}{\begin{equation}} \nc{\eq}{\end{equation}} \nc{\bZ}{{\bf Z}}
\nc{\ba}{\left(\begin{array}{cc}}  \nc{\ea}{\end{array}\right)}
\nc{\bR}{{\bf R}}
\nc{\Dl}{\Delta} \nc{\bT}{{\bf T}} \nc{\bC}{{\bf C}}
\nc{\la}{\lambda} \nc{\cHp}{{\cal H^{\perp}}}
\nc{\al}{\alpha} \renewcommand{\sp}{\vspace{1ex}}
\nc{\noi}{\noindent}\nc{\hff}{\scriptstyle{\hf}} \nc{\si}{\sigma}
\nc{\pl}{\partial} \nc{\iy}{\infty} \nc{\ps}{\psi} \nc{\et}{\eta}
\nc{\ch}{\raisebox{.4ex}{$\chi$}} \nc{\cd}{\cdots} \nc{\tn}{\otimes}
\nc{\cD}{\cal D} \nc{\cH}{{\cal H}} \nc{\inv}{^{-1}} \nc{\cN}{\cal N}
\nc{\tl}{\tilde} \nc{\nm}{\parallel} \nc{\sa}{^{(\al)}}
\nc{\onetwo}[2]{\left(\begin{array}{cc}#1&#2\end{array}\right)}
\nc{\twotwo}[4]{\left(\begin{array}{cc}#1&#2\\&\\#3&#4\end{array}\right)}
\nc{\twoone}[2]{\left(\begin{array}{c}#1\\\\#2\end{array}\right)}
\nc{\sgn}{{\rm sgn}} \nc{\Ga}{\Gamma}

\begin{document} \begin{center}{\large\bf On the Relation
Between
Orthogonal, Symplectic\\  \vspace{.7ex} and Unitary Matrix
Ensembles}\end{center}

\sp\begin{center}{{\bf Harold Widom}\\
{\it Department of Mathematics\\
University of California, Santa Cruz, CA 95064, USA\\
e-mail address: widom@math.ucsc.edu}}\end{center}\sp

\begin{abstract}

For the unitary ensembles of $N\times N$ Hermitian matrices associated
with a
weight function $w$ there is a kernel, expressible in terms of the
polynomials orthogonal with respect to the weight function, which plays
an important role. For
the orthogonal and symplectic ensembles of Hermitian matrices there are
$2\times2$ matrix kernels, usually constructed  using skew-orthogonal
polynomials, which play an analogous role. These matrix kernels are
determined by their upper left-hand entries. We derive formulas
expressing these
entries in terms of the scalar kernel for the corresponding  unitary
ensembles. We also show that whenever $w'/w$ is a rational
function the entries are equal to the scalar kernel plus
some extra terms whose number equals the order of $w'/w$. General
formulas are obtained for these extra terms. We do not use
skew-orthogonal polynomials in the derivations.
\end{abstract}\sp

\renewcommand{\theequation}{1.\arabic{equation}}
\begin{center}{\bf 1. Introduction}\end{center}

In the most common ensembles of $N\times N$ Hermitian matrices the
probability density $P_N(x_1,\cd,x_N)$ that the eigenvalues lie in
infinitesimal neighborhoods of $x_1,\cd,x_N$ is given by
\[P_N(x_1,\cd,x_N)=c_N\prod_{j<k}|x_j-x_k|^{\be}\,\prod_jw(x_j),\] where
$\be=1,\ 2$ or 4 (corresponding to the orthogonal, unitary and
symplectic
ensembles, respectively), $w(x)$ is a weight function and $c_N$ is a
normalization constant.

For the unitary matrix ensembles an important role is played by the
kernel
\bq K_N(x,y)=\sum_{k=0}^{N-1}\ph_k(x)\,\ph_k(y),\label{K}\eq where
$\{\ph_k(x)\}$ is the sequence obtined by orthonormalizing the sequence
$\{x^k\,w(x)^{1/2}\}$. The probability density is expressed in terms of
it by
\[P_N(x_1,\cd,x_N)={1\ov N!}\,\det\,(K_N(x_j,x_k))_{j,k=1,\cd,N}.\] More
generally
the $n$-point correlation function $R_n(x_1,\cd,x_n)$, the probability
density that $n$ of the eigenvalues, irrespective of order, lie in
infinitesimal
neighborhoods of
$x_1,\cd,x_n$, is given by the formula \bq
R_n(x_1,\cd,x_n)=\det\,(K_N(x_j,x_k))_{j,k=1,\cd,n}.\label{R}\eq And the
probability $E(0;J)$ that the set $J$ contains no eigenvalues is equal
to the
Fredholm determinant of the kernel $K_N(x,y)\,\ch_J(y)$, where $\ch$
denotes
characteristic function.

For the orthogonal and symplectic ensembles there are $2\times 2$ matrix
kernels which play analogous roles. In this case the determinant in
(\ref{R})
is to be interpreted as a quaternion determinant (it is a linear
combination
of traces of products of matrix entries of the block matrix on the right
side), and  the square of $E(0;J)$ equals the Fredholm determinant of
the
matrix kernel.  (The last fact can be deduced from the computation in
\cite{M1}, sec. A.7. A direct derivation is given in \cite{TW2}.) In the
case
of the orthogonal ensembles we shall always assume that $N$ is even. The
kernels for the orthogonal and  symplectic ensembles are of the form \bq
K_{N1}(x,y)=\twotwo{S_{N1}(x,y)}{S_{N1}D(x,y)}{IS_{N1}(x,y)-\ep(x-y)}{S_{N1}(y,x)}
\label{1K}\eq and \bq K_{N4}(x,y)=\hf\twotwo
{S_{N4}(x,y)}{S_{N4}D(x,y)}{IS_{N4}(x,y)}{S_{N4}(y,x)} \label{4K}\eq
respectively. Here $\ep(x)={\hff}\,\sgn(x)$ and the explanation for the
notation is this: the $S_{N\be}(x,y)$ are certain sums of products and
if $S_{N\be}$ is the operator with kernel  $S_{N\be}(x,y)$ then
$S_{N\be}D(x,y)$ is the kernel of $S_{N\be}D$ ($D=$ differentiation) and
$IS_{N\be}(x,y)$ is
the kernel of $IS_{N\be}$ ($I=$ integration, more or less). We shall
write these out below. One can see from this description that once the
kernels
$S_{N\be}(x,y)$ are known then so are the others.

Matrix kernels were first introduced by Dyson \cite{D} for his circular
ensembles and he established the analogue of formula (\ref{R}) for
the correlation functions. Later, Mehta \cite{M1} and Mehta  and
Mahoux
\cite{MM} found matrix kernels for the ensembles  of Hermitian matrices,
and
expressed them in terms of systems of skew-orthogonsal polynomials.
These are
like orthogonal polynomials but the inner product (different in the
$\be=1$
and $\be=4$ cases) is antisymmetric instead of symmetric. In terms of
them
one obtains for $S_{N\be}(x,y)$ sums like the one in $(\ref{K})$ but
which are a little more complicated. A problem here is that
the skew-orthogonal polynomials are not always that easy to compute and,
even if they are, the sums involving them may not be easy to handle. For
example, one is often interested in  scaling limits as $N\ra\iy$ and in
order to
do this it helps to have a good representation for the sum.

In this paper we shall not use skew-orthogonal polynomials at all.
Instead, we
shall use the expressions for the various matrix kernels in the
general form given in \cite{TW2}, and derive general formulas for the
$S_{N\be}(x,y)$
in terms of the scalar kernel $K_N(x,y)$ given by (\ref{K}), with $N$
replaced by $2N$ when $\be=4$ and $w$ replaced by $w^2$ when $\be=1$.
More exactly, we shall express the
operators whose kernels are the $S_{N\be}(x,y)$ in terms of the operator
whose kernel is $K_N(x,y)$. These are given in Theorem 1 below. 

The formulas can be brought to a very concrete form whenever the support
$\cD$ of $w$
is a finite union of finite or infinite intervals and $w'/w$ is equal to
a rational
function on $\cD$. (Such weight functions are called {\it
semi-classical} since
they include the weight functions for all the classical orthogonal
polynomials.)
We find then that the $S_{N\be}(x,y)$
are equal to the appropriate scalar kernel $K_N(x,y)$ plus
some extra terms whose number is independent of $N$. This number
equals
the order of $w'/w$, the sum of the orders of its poles in the extended
complex plane. We must also count as a simple pole any end-point of
$\cD$ where $w'/w$ is
analytic. Thus, for the Gaussian ensembles ($w(x)=e^{-x^2}$) and
Laguerre ensembles ($w(x)=x^{\al}\,e^{-x}$) there will  one extra term
because of the simple poles at $\iy$ and 0 respectively, and for the
Jacobi ensemble ($w(x)=(1-x)^{\al}(1+x)^{\be}$) there will be two extra
terms
because of the simple poles at $\pm 1$.
For the Legendre  ensemble on $(-1,\,1)$ there will also be two extra
terms
although $w'/w=0$ in this case.

We shall produce explicit formulas for the extra terms, which are
given in
Theorem 2. These will be used to work out the cases of
the Gaussian ensembles (well-known \cite {M2}) and the Laguerre
ensembles (known apparently only in the case $\al=0$ \cite{NW}).

To apply our formulas to the Laguerre
ensemble we require at first that $\al>0$ so that Theorem 1 is 
applicable. The formulas for general $\al>-1$ are then obtained by analytic
continuation. Similar analytic continuation arguments apply quite
generally. (See the remark at the end of section 3.) For example, for the Legendre 
ensemble we would start with the formulas
for the Jacobi ensemble for $\al,\ \be>0$ and then take the analytic
continuation (or limit) to obtain the fomulas for $\al=\be=0$. This is
the reason the end-points $\pm 1$ count as poles.

The recent announcement \cite{SV} has some elements in common with ours. A
generalization of the Laguerre ensemble was considered there where
$e^{-x}$ was replaced by the exponential of an arbitrary polynomial and the
occurrence of
only finitely many extra terms was established, without their being
evaluated, using skew-orthogonal polynomials. This fact was used to
deduce universality for this class of ensembles.\sp

\begin{center}{\bf 2. The general identities}\end{center}
\setcounter{equation}{0}\renewcommand{\theequation}{2.\arabic{equation}}

We start with the expressions for the various matrix kernels in
the form given in \cite{TW2}. (The notation here is slightly different.)
Taking the symplectic enembles first, we let $\{p_j(x)\}$ be any
sequence of
polynomials of exact degree $j$ and define
$\ph_j(x)=p_j(x)\,w(x)^{1/2}$. Let
$M$ be the $2N\times2N$ matrix with $j,k$ entry ($j,k=0,\cd,2N-1$)
\bq m_{jk}=\hf\int(p_j(x)\,p_k'(x)-p_j'(x)\,p_k(x))\,w(x)\,dx
=\hf\int(\ph_j(x)\,\ph_k'(x)-\ph_j'(x)\,\ph_k(x))\,dx.\label{M4}\eq
This matrix
is invertible and we write $M^{-1}=(\mu_{jk})$. Then \bq
S_{N4}(x,y)=\sum_{j,k=0}^{2N-1}
\ph_j'(x)\,\mu_{jk}\,\ph_k(y)\label{N4}\eq
and \[IS_{N4}(x,y)=\sum_{j,k=0}^{2N-1} \ph_j(x)\,\mu_{jk}\,\ph_k(y), \
\ \ S_{N4}D(x,y)=-\sum_{j,k=0}^{2N-1} \ph_j'(x)\,\mu_{jk}\,\ph_k'(y).\]
Any family of  polynomials leads to the same matrix kernel. Of course at
this point the formulas look quite bad because of the $\mu_{jk}$.

For the orthogonal ensembles we take we take the $p_j$ as before but
this time define
$\ph_j(x)=p_j(x)\,w(x)$ and let $M$ be the $N\times N$ matrix
with $j,k$ entry ($i,j=0,\cd,N-1$)
\bq m_{jk}=\int\int\ep(x-y)\,p_j(x)\,p_k(y)\,w(x)\,w(y)\,dy\,dx
=\int\ph_j(x)\;\,\ep\ph_k(x)\,dx.\label{M1}\eq
Here $\ep$ denotes the operator with kernel $\ep(x-y)$.
Again $M$ is invertible, we write $M^{-1}=(\mu_{jk})$, and the formulas
for the kernels are
\[S_{N1}(x,y)=-\sum_{j,k=0}^{N-1}\ph_j(x)\,\mu_{jk}\,\ep\ph_k(y),\]
\[IS_{N1}(x,y)=-\sum_{j,k=0}^{N-1}\ep\ph_j(x)\,\mu_{jk}\,\ep\ph_k(y),
 \ \ S_{N1}D(x,y)=\sum_{j,k=0}^{N-1}\ph_j(x)\,\mu_{jk}\,\,\ph_k(y).\]

We change notation so that we can treat the two cases at the
same time---we shall see that they are interrelated. We continue
to use the notations $N$ and $w$, but when $\be=4$ the $N$ here will be
the $2N$
of (\ref{N4}) and when $\be=1$ the $w$ here will be square of the weight
function  in (\ref{M1}). Thus in both cases $N$ is even, we take $p_j$
to be polynomials of exact degree $j$  and set
$\ph_j=p_j\,w^{1/2}$. The matrices $(m^{(\be)}_{jk})$ and
$(\mu^{(\be)}_{jk})$ are the $M$ and $M\inv$ corresponding to the
$\be=4$ and 1 ensembles. We set
\[S^{(4)}_N(x,y)=\sum_{j,k=0}^{N-1}\ph_j'(x)\,\mu^{(4)}_{jk}\,\ph_k(y),
\ \ \ S^{(1)}_N(x,y)=-\sum_{j,k=0}^{N-1}\ph_j(x)\,\mu^{(1)}_{jk}\,\ep\ph_k(y).\]
Finally, $K_N(x,y)$ will denote the $\be=2$ scalar kernel (\ref{K}).

We denote by $\cH$ be the linear space spanned by
the functions $\ph_0,\cd,\ph_{N-1}$, in other words the set of all
functions of the form $w^{1/2}$ times a polynomial of degree less than
$N$. We denote by $K$ be the projection operator onto $\cH$. Its kernel is
$K_N(x,y)$. Finally, we denote by $S^{(4)}$ the operator with kernel
$S^{(4)}_N(x,y)$ and by ${S^{(1)}}'$ the operator with kernel
$S^{(1)}_N(y,x)$.

The following lemma will identify these operators. We think of our weight
functions as defined on all of \bR, and our basic assumptions are 
\bq \cH\subset L^1(\bR),\ \ \ \ D\cH\subset L^1(\bR). \label{assume}\eq
The former is needed even to define the ensembles. The latter is restrictive 
and implies in particular that all the $\ph_k$ are
absolutely continuous.  We use the notations $D_\cH$ and
$\ep_\cH$ for the restrictions of the operators $D$ and $\ep$,
respectively, to $\cH$.\sp

\noi{\bf Lemma}. The operators $KD_\cH$ and $K\ep_\cH$ are invertible
and
\[S^{(4)}|_\cH=D(KD_\cH)\inv,\ \ \  S^{(4)}|_\cHp=0,\]
\[{S^{(1)}}'|_\cH=\ep(K\ep_\cH)\inv,\ \ \ {S^{(1)}}'|_\cHp=0.\]\sp

\noi{\bf Proof}. Integrating by parts the second integral in (\ref{M4})
shows that $m^{(4)}_{jk}=\int\ph_j(x)\,\ph_k'(x)\,dx$. (This is where the absolute 
continuity of the $\ph_k$ come in.) Thus for $i=0,\cd,N-1$,
\[S^{(4)}K\ph_i'=\sum_{j,k}\ph_j'\,\mu^{(4)}_{jk}\,(\ph_k,\,\ph_i')
=\sum_{j,k}\ph_j'\,\mu^{(4)}_{jk}\,m^{(4)}_{ki}=\sum_j\ph_j'\,\dl_{ji}=\ph_i'.\]
Since the $\ph_i$ span $\cH$ we see that $S^{(4)}\,K\,\ph'=\ph'$ for all
$\ph\in\cH$.
This shows that
$KD_\cH$ is a one-one, and hence invertible, operator on $\cH$, and also
that
$S^{(4)}|_\cH=D(KD_\cH)\inv$. Of course $S^{(4)}|_\cHp=0$ since each
$\ph_k\in\cH$. This proves the first part of the lemma. For the second,
observe that
by the antisymmetry of $(m^{(1)}_{jk})$ the formula for $S^{(1)}_N(y,x)$
can be obtained from the formula for $S^{(4)}_N(x,y)$ by replacing $m^{(4)}_{jk}$
by $m^{(1)}_{jk}$ and $\ph_j'(x)$ by $\ep\ph_j(x)$.
Thus the  second part of the lemma can be proved just as the first,
replacing $D$ everywhere by $\ep$.\sp

To identify $(KD_\cH)\inv$ and  $(K\ep_\cH)\inv$ more concretely we
shall enlarge
the domains of $D$ and $\ep$. We have $D\cH\subset L^1(\bR)$ by
assumption, and
$\ep\cH\subset L^{\iy}(\bR)$ since $\cH\subset L^1(\bR)$. It is easy to
see
that the operators
\[D:\cH+\ep\cH\ra \cH+D\cH,\ \ \ \ep:\cH+D\cH\ra \cH+\ep\cH\]
are mutual inverses. In the following, $I_{\cH+D\cH}$ and
$I_{\cH+\ep\cH}$ will
denote the identity operators on the spaces $\cH+D\cH$ and $\cH+\ep\cH$,
respectively.
\sp

\noi{\bf Theorem 1}. We have
\bq S^{(4)}=(I_{\cH+D\cH}-(I-K)DK\ep)\inv K,\label{S4}\eq
\bq {S^{(1)}}'=(I_{\cH+\ep\cH}-(I-K)\ep KD)\inv K.\label{S1}\eq

\noi{\bf Proof}. Since $D$ and $\ep$ are mutual inverses we
might guess that a good approximation to the inverse of $KD_\cH$ is
$K\ep_\cH$, where $\ep_\cH$ denotes the restriction of $\ep$ to $\cH$.
With this in view, we compute
\[K\ep KD_\cH=K\ep D_\cH-K\ep(I-K)D_\cH=I_{\cH}-K\ep(I-K)D_\cH,\]
where $I_\cH$ denotes the
identity operator on $\cH$. The operator on the right side is invertible
since both $KD_\cH$ and $K\ep_\cH$ are, and we deduce that
\[ (KD_\cH)\inv=(I_\cH-K\ep(I-K)D_\cH)\inv K\ep_\cH.\]
Hence by the lemma,
\[S^{(4)}|_\cH=D_{\cH}(KD_\cH)\inv=KD_{\cH}(KD_\cH)\inv+(I-K)D_\cH(KD_\cH)\inv\]
\[=I_\cH+(I-K)D_\cH(I_\cH-K\ep(I-K)D_\cH)\inv K\ep_\cH.\]
Recall that the domain of $\ep$ is $\cH+D\cH$ and set
\[A=(I-K)D_\cH:\cH\ra \cH+D\cH,\ \ \ B=K\ep:\cH+D\cH\ra\cH.\]
Then
\[I_{\cH+D\cH}+(I-K)D_\cH(I_\cH-K\ep(I-K)D_\cH)\inv K\ep\]
is equal in this notation to
$I_{\cH+D\cH}+A(I_\cH-BA)\inv B$. This in turn equals
$(I_{\cH+D\cH}-AB)\inv.$ Hence restricting to $\cH$ gives
\[S^{(4)}|_\cH=(I_{\cH+D\cH}-(I-K)DK\ep)\inv\Big|_\cH.\]
Since $S^{(4)}|_\cHp=K|_\cHp=0$
this gives (\ref{S4}), and (\ref{S1}) is obtained by an analogous
argument, interchanging the roles of $D$ and $\ep$.\sp

\noi{\bf Remark}. The identities of the theorem
may be restated in the rather more complicated form
\[ S^{(4)}=K+(I-K)DK\ep(I_{\cH+D\cH}-(I-K)DK\ep)\inv K,\]
\[ {S^{(1)}}'=K+(I-K)\ep KD(I_{\cH+\ep\cH}-(I-K)\ep KD)\inv K.\]
The summands on the right may be thought of as corrections and we see
that
they will be of finite rank (independent of
$N$) whevever $(I-K)DK\ep$ and $(I-K)\ep KD$ are. This will be true
whenever the commutator $[D,\,K]$ is, which will be the case in what
follows.

\begin{center}{\bf 3. The case of rational}\ {\boldmath
$w'/w$}\end{center}
\setcounter{equation}{0}\renewcommand{\theequation}{3.\arabic{equation}}

We assume now that $w'/w$ is a rational function on the support of $w$
and, at first, that (\ref{assume}) holds so that Theorem 1 is applicable.
We explain at the end of this section how to remove the restriction 
in the cases of greatest interest.
From now on it will be convenient to take the $p_j$ to be the
polynomials orthonormal with respect to the weight function $w$ so that
the $\ph_j$ are orthonormal with respect to Lebesgue measure.

It follows from the Christoffel-Darboux formula that there is a
representation
\[K_N(x,y)=a_N\,{\ph_N(x)\,\ph_{N-1}(y)-\ph_{N-1}(x)\,\ph_N(y)\ov
x-y}\]
\bq=a_N\,\onetwo{\ph_N(x)}{\ph_{N-1}(x)}\twotwo{0}{1}{-1}{0}
\twoone{\ph_N(y)}{\ph_{N-1}(y)}{\mbox \Large/}(x-y)  \label{Krep}\eq
for a certain constant $a_N$. This holds for an arbitrary weight function.
Whenever $w'/w$ is a rational function there is a differentiation formula
\[\twoone{\ph_N'}{\ph_{N-1}'}=\twotwo{A}{B}{-C}{-A}\twoone{\ph_N}{\ph_{N-1}}\]
where $A(x),\ B(x)$ and
$C(x)$ are rational functions  whose poles are among those of $w'/w$,
counting multiplicity. (See \cite{TW1}, sec. 6.) From this
and (\ref{Krep}) we find that the kernel of $[D,\,K]$, which equals
$(\pl_x+\pl_y)\,K_N(x,y)$, is equal to
\bq a_N\,\onetwo{\ph_N(x)}{\ph_{N-1}(x)} \twotwo{\displaystyle{{C(x)-C(y)\ov
x-y}}}{\displaystyle{{A(x)-A(y)\ov x-y}}} {\displaystyle{{A(x)-A(y)\ov
x-y}}}{\displaystyle{{B(x)-B(y)\ov x-y}}}\twoone{\ph_N(y)}{\ph_{N-1}(y)}.\label{D,K}\eq
It follows from this that
$[D,\,K]$ is a finite rank operator, and that its kernel is expressible
in terms of the functions
\bq x^k\,\ph_{N-1}(x),\ \ \ x^k\,\ph_N(x),\ \ (0\leq k<n_{\iy})\label{phinf}\eq
where $n_{\iy}$ is the order of $w'/w$ at infinity and, for each finite
pole $x_i$ of $w'/w$, the functions
\bq (x-x_i)^{-k-1}\,\ph_{N-1}(x),\ \ \ (x-x_i)^{-k-1}\,\ph_N(x),\ \ (0\leq k<n_{x_i}).
\label{phfin}\eq
where $n_{x_i}$ is the order of $w'/w$ at $x_i$. This is seen by
expanding the functions appearing in the central matrix in (\ref{D,K}), which is
simple algebra.

In the space spanned by these $2n$ functions ($n$ is
the total order of $w'/w$) there is a subspace of dimension $n$ contained in
$\cH$ and a subspace of dimension $n$ contained in $\cHp$. To see the first,
an inductive argument using the three-term recurrence formula shows that the subspace
spanned by the funcions (\ref{phinf}) contains the $n_{\iy}$ functions
$\ph_{N-k}\ (0<k\leq n_{\iy})$ which lie in $\cH$. The functions (\ref{phfin}) span a space
of dimension $2\sum n_{x_i}$ consisting of functions which equal $w^{1/2}$ times
rational functions which may have poles at the $x_i$ of order
$n_{x_i}$. A function in this space will belong to $\cH$ if the
principal parts at all these poles vanish. This gives $\sum n_{x_i}$ conditions
in a space of dimension $2\sum n_{x_i}$, giving us a subspace of
dimension $\sum n_{x_i}$ which is contained in $\cH$. Thus the space spanned by
the functions (\ref{phinf}) and (\ref{phfin}) together contains a subspace of
dimension $n$ contained in $\cH$. To see that there is an $n$-dimensional subspace
lying entirely in $\cH^{\perp}$, observe that $\cH$ is spanned by the functions
\[\ph_{N-k}\ (k<n_{\iy}),\ \ \ \ph_k\ (k<n-n_{\iy}),\ \ \
\prod(x-x_i)^{n_{x_i}}\,x^k\ (k<N-n).\]
Our $2n$ functions are all orthogonal to the last of these, whereas
orthogonality to the remaining ones imposes $n$ conditions, giving a subspace of
dimension $n$ which is contained in $\cHp$.

It follows from the preceding discussion that the space spanned by the functions
(\ref{phinf}) and (\ref{phfin}) contains $n$ linearly independent functions
$\ps_1,\cd,\ps_n$ lying in $\cH$ and $n$ linearly independent functions
$\ps_{n+1},\cd,\ps_{2n}$ lying in $\cHp$. And we have a representation
\bq [D,\,K]=\sum_{i,j=1}^{2n} A_{ij}\,\ps_i\tn\ps_j\label{Dcom}\eq
for some constants $A_{ij}$ which can be determined from (\ref{D,K})
once we have fixed the $\ps_i$. (We use the
notation $a\tn b$ for the operator with kernel $a(x)\,b(y)$.) It follows
that also
\bq [\ep,\,K]=\sum_{i,j=1}^{2n}A_{ij}\,\ep\ps_i\tn\ep\ps_j.\label{epcom}\eq
Here we used $\ep D=D\ep=I$, the antisymmetry of $\ep$ and the easy fact that
$(a\tn b)\,T=a\tn(T'b)$
for any operator $T$. These will be used again below without comment.

The matrix $A=(A_{ij})$ is symmetric since $K$ is symmetric and $D$ is antisymmetric.
(We hope this $A$ will not be confused with the function $A$ appearing in (\ref{D,K}).)
Since $K$ is the projection operator onto $\cH$ the commutator
$[D,\,K]$ takes $\cH$ to $\cHp$ and $\cHp$ to 0. Hence
\bq A_{ij}=0\ {\rm if}\ i,j\leq n\ {\rm or}\ i,j>n.\label{A}\eq

After a little more notation we shall be able to state the formulas.
We already have the matrix $A$. We define the matrix $B$ by
\[B_{ij}=(\ep\ps_i,\,\ps_j).\]
Define $J$ to be the matrix whoese $i,j$ entry equals 1 if $i=j\leq n$ and 0
otherwise. Finally, set
\[C=J+BA\]
and write $A_0$ for the matrix obtained from $A$ by deleting its last $n$ columns,
$C_0$ for the matrix obtained from $C$ by deleting its last $n$ rows and $C_{00}$
for the matrix obtained from $C$ by deleting its last $n$ rows and its last $n$
columns. Observe that by (\ref{A}) the first $n$ rows of $A_0$ are zero.\sp

\noi{\bf Theorem 2}. We have
\bq S^{(4)}_N(x,y)=K_N(x,y)-\sum_{i>n,\;j}(A_0\,C_{00}\inv\,C_0)_{ij}\;\ps_i(x)\;\ep\ps_j(y),
\label{S4rat}\eq
\bq S^{(1)}_N(x,y)=K_N(x,y)-\sum_{i\leq n,\;j}[AC(I-BAC)\inv]_{ji}\;\ps_i(x)\;\ep\ps_j(y).
\label{S1rat}\eq

\noi{\bf Proof}. Using (\ref{Dcom}) we find
\[(I-K)DK\ep=[D,\,K]K\ep=(\sum_{i,j}A_{ij}\,\ps_i\tn\ps_j)\,K\ep
=-\sum_{j\leq n,\;i}A_{ij}\,\ps_i\tn\ep\ps_j\]
since $K\ps_j=\ps_j$ when $j\leq n$ and $K\ps_j=0$ when $j>n$. Thus
\[I-(I-K)DK\ep=I+\sum_{j\leq n,\;i}A_{ij}\,\ps_i\tn\ep\ps_j.\]
Now if we have a finite rank operator $\sum a_i\tn b_i$
then
\bq (I+\sum a_i\tn b_i)\inv=I-\sum_{i,j}\,T_{\;ij}\inv\,a_i\tn b_j,\label{abinv}\eq
where $T$ is the matrix with entries
\[ T_{ij}=\dl_{ij}+(b_i,\,a_j).\]
In our case $i,\,j\leq n$ and
\[a_i=\sum_kA_{ki}\,\ps_k,\ \ \ b_i=\ep\ps_i,\]
so
\[T_{ij}=\dl_{ij}+\sum_k(\ep\ps_i,\,\ps_k)\,A_{kj}=\dl_{ij}+\sum_kB_{ik}\,A_{kj}.\]
This equals $(I+BA)_{ij}=C_{ij}$ and so we have shown
\[(I-(I-K)DK\ep)\inv=I-\sum_{i,j\leq n}(C_{00})\inv_{\;ij}\,
(\sum_kA_{ki}\,\ps_k\tn\ep\ps_j),\]
whence
\bq S^{(4)}=(I-(I-K)DK\ep)\inv K=K-\sum_{i,j\leq n}\sum_kA_{ki}\,(C_{00})\inv_{\;ij}\,
\ps_k\tn K\ep\ps_j.\label{S4temp}\eq
To compute $K\ep\ps_j$ we apply (\ref{epcom}) to $\ps_j$, using the fact
that $\ps_j\in\cH$, to obtain
\[ K\ep\ps_j=\ep\ps_j-\sum_{l,k}A_{lk}\,\ep\ps_l\,(\ep\ps_k,\,\ps_j)=
\ep\ps_j-\sum_{l,k}A_{lk}\,B_{kj}\,\ep\ps_l\]
\bq=\ep\ps_j+\sum_l(BA)_{jl}\ep\ps_l=\sum_l C_{jl}\ep\ps_l.\label{Keps}\eq
Here we used the symmetry of $A$ and the antisymmetry of $B$. Substituting this into
(\ref{S4temp}) gives
\[S^{(4)}=(I-(I-K)DK\ep)\inv K=K-\sum_{k,\,l}(A_0\,C_{00}\inv\,C_0)_{kl}\,\ps_k
\tn \ep\ps_l,\]
which is the same as (\ref{S4rat}).

To derive (\ref{S1rat}) we use (\ref{epcom}) and find that
\bq (I-K)\ep KD=[\ep,\,K]KD=-\sum_{i,j}A_{ij}\,\ep\ps_i\tn DK\ep\ps_j.\label{epKKD}\eq
Using (\ref{Dcom}) again and the fact that $D\ep=I$ we see that
\[DK\ep\ps_j=K\ps_j+\sum_{k,l}A_{kl}\,\ps_k\,(\ps_l,\,\ep\ps_j)=K\ps_j+\sum_{k,l}B_{jl}\,
A_{lk}\ps_k.\]
Again we use the fact that $K\ps_j=\ps_j$ when $j\leq n$ and $K\ps_j=0$ when $j>n$.
If we recall the definitions of $J$ and $C$ we see that we have shown
$DK\ep\ps_j=\sum_k C_{jk}\,
\ps_k$. Substituting this into (\ref{epKKD}) gives
\[(I-K)\ep KD=-\sum_{i,j,k}A_{ij}\,C_{jk}\,\ep\ps_i\tn \ps_k\]
and so
\[I-(I-K)\ep KD=I+\sum_{i,k}(AC)_{ik}\,\ep\ps_i\tn \ps_k.\]
We use (\ref{abinv}) again, this time with $i,j\leq 2n$ and
\[a_i=\sum_{k}(AC)_{ki}\,\ep\ps_k,\ \ \ b_i=\ps_i.\]
Now we have
\[T_{ij}=\dl_{ij}+\sum_k(\ps_i,\,\ep\ps_k)\,(AC)_{kj}=
\dl_{ij}-\sum_kB_{ik}\,(AC)_{kj}=(I-BAC)_{ij}.\]
Hence  (\ref{abinv}) gives
\[(I-(I-K)\ep KD)\inv=\sum_{i,j}(I-BAC)\inv_{ij}\sum_k(AC)_{ki}\,\ep\ps_k\tn\ps_j
=\sum_{j,k}[AC(I-BAC)\inv]_{kj}\,\ep\ps_k\tn\ps_j.\]
To obtain ${S^{(1)}}'$ we must right-multiply by $K$, which has the effect of
imposing the restriction $j\leq n$. After taking transposes and changing notation
we obtain (\ref{S1rat}).\sp

\noi{\bf Remark}. Here is how to extend the results to the case where 
the second part of (\ref{assume})
may not be satisfied but the support $\cD$ of $w$ consists of a finite union of
intervals. Denote now by $x_i$ the poles of $w'/w$ together with all finite end-points
of $\cD$ where $w'/w$ is analytic. Then we can write
\[w(x)=\prod_i(x-x_i)^{\al_i}\,w_0(x),\]
where $w_0$ satisfies (\ref{assume}) and each $\al_i>-1$. Think of $w$, and therefore 
the kernels $K_N(x,y)$, $S^{(4)}_N(x,y)$ and $S^{(1)}_N(x,y)$, as functions of the
$\al_i$. Theorem 2 would apply to $w$ itself if all the $\al_i>0$ since then
(\ref{assume}) would be satisfied. But the constituents of these kernels are real-analytic
functions of the $\al_i$, so the formulas for $\al_i>-1$ (and therefore for our
given weight function $w$) can be obtained by analytic continuation of the formulas 
for $\al_i>0$. 

\begin{center}{\bf 4. The Gaussian and Laguerre ensembles}\end{center}
\setcounter{equation}{0}\renewcommand{\theequation}{4.\arabic{equation}}

These are (essentially the only) cases where $n=1$ and are
especially simple, as we shall now see. 

By the symmetry of $A$ and (\ref{A}), $A$ has the form
\[A=\twotwo{0}{\la}{\la}{0}\]
for some constant $\la$. There arise two functions, $\ps_1$ and $\ps_2$,
the first lying in $\cH$ and the second lying in $\cH^{\perp}$.

Since the two $2\times 2$ matrices $A$ and $B$ have 0 diagonal entries, $AB$ is
a diagonal matrix and therefore so is $C$. Therefore (\ref{Keps}), in which $j=1$,
says that $K\ep\ps_1=C_{11}\,\ps_1$. Since $K\ep_\cH$ is invertible $C_{11}\neq0$, and
so $\ep\ps_1\in\cH$. This implies that all entries of $B$ vanish, so $B=0,\ C=J$.
It is immediate from these facts that
\[A_0\,C_{00}\inv\,C_0=AC(I-BAC)\inv=\twotwo{0}{0}{\la}{0}.\]
Hence by Theorem 2,
\[ S^{(4)}_N(x,y)=K_N(x,y)-\la\,\ps_2(x)\;\ep\ps_1(y),\]
\[S^{(1)}_N(x,y)=K_N(x,y)-\la\,\ps_1(x)\;\ep\ps_2(y).\]
It remains to find the constant $\la$ and the functions $\ps_1$ and
$\ps_2$ in the two cases.\sp

\noi{\bf The Gaussian ensembles}.  Here the pole is at $x=\iy$. Clearly
$\ps_1=\ph_{N-1}$ and $\ps_2=\ph_N$
in this case. Moreover we have for this ensemble
\[a_N=\sqrt{N/2},\ \ \ A(x)=-x,\ \ \ B(x)=C(x)=\sqrt{2N},\] so we find
from
(\ref{D,K}) that
\[A=\twotwo{0}{-\sqrt{N/2}}{-\sqrt{N/2}}{0},\]
which gives $\la=-\sqrt{N/2}$. Therefore
\[ S^{(4)}_N(x,y)=K_N(x,y)+\sqrt{N/2}\;\ph_{N}(x)\;\ep\ph_{N-1}(y),\]
\[ S^{(1)}_N(x,y)=K_N(x,y)+\sqrt{N/2}\;\ph_{N-1}(x)\;\ep\ph_N(y).\]\sp

\noi{\bf The Laguerre ensembles}. Here $w(x)=x^{\al}\,e^{-x}$ and
\bq p_j(x)=\sqrt{{j!\ov\Gamma(j+\al+1)}}L_j\sa(x),\label{Lag}\eq
where $L_j\sa$ is the generalized Laguerre polynomial. We assume at first
that $\al>0$ so that (\ref{assume}) holds. In the notation  of (\ref{Krep}) and
(\ref{D,K}) we have
\[a_N=-\sqrt{N(N+\al)},\ \ \ A(x)=x\inv(N+{\al\ov2})-\hf,\ \ \
B(x)=C(x)=-x\inv\sqrt{N(N+\al)}.\]
The pole is at $x=0$, and if we set $\xi_j(x)=x\inv\ph_j(x)$
then (\ref{D,K}) becomes
\bq\sqrt{N(N+\al)}\,\onetwo{\xi_N(x)}{\xi_{N-1}(x)}
\twotwo{-\sqrt{N(N+\al)}}{N+\displaystyle{\al\ov2}}
{N+\displaystyle{\al\ov2}}{-\sqrt{N(N+\al)}}
\twoone{\xi_N(y)}{\xi_{N-1}(y)}.\label{D,KLag}\eq

Our functions $\ps_1$ and $\ps_2$ are linear combinations of $\xi_N$ and $\xi_{N-1}$,
with $\ps_1$ lying in $\cH$ and $\ps_2$ lying in $\cHp$. Clearly $\ps_1$ is a constant
times $p_{N-1}(0)\,\xi_N(x)-p_N(0)\,\xi_{N-1}(x)$. Since
$L_N\sa(0)/L_{N-1}\sa(0)=(N+\al)/ N$
we see using (\ref{Lag}) that we may take
\[\ps_1=\sqrt{N}\,\xi_N-\sqrt{N+\al}\,\xi_{N-1}.\]
For $\ps_2$, it follows from the discussion near the beginning of the last
section that the appropriate linear combination of $\xi_N$ and $\xi_{N-1}$ may be
found by requiring that it be orthogonal to $\ph_0$. From the fact that
$\int_0^{\iy}L_m\sa(x)\,x^{\al-1}\,e^{-x}\,dx=\Ga(\al)$ 
and from (\ref{Lag}) we see that the linear combination
\[\ps_2=\sqrt{N+\al}\,\xi_N-\sqrt{N}\,\xi_{N-1}\]
does the job.

Solving for $\xi_N$ and $\xi_{N-1}$ in terms of $\ps_1$ and $\ps_2$ and
substituting into (\ref{D,KLag}) we obtain for the kernel of
$[D,\,K]$,
\[-{\sqrt{N(N+\al)}\ov2}\,\onetwo{\ps_1(x)}{\ps_2(x)}\, \twotwo{0}{1}{1}{0}\,
\twoone{\ps_1(y)}{\ps_2(y)}.\]
Therefore
\[\la=-{\sqrt{N(N+\al)}\ov2}.\]
Hence for this ensemble we find that
$S^{(4)}_N(x,y)$ is equal to $K_N(x,y)$ plus
\bq{\sqrt{N(N+\al)}\ov2}\,(\sqrt{N+\al}\;\xi_N(x)-\sqrt{N}\;\xi_{N-1}(x))\;
(\sqrt{N}\;\ep\xi_N(y)-\sqrt{N+\al}\;\ep\xi_{N-1}(y))\label{Lag4}\eq and
that
$S^{(1)}_N(x,y)$ is equal to $K_N(x,y)$ plus \bq{\sqrt{N(N+\al)}\ov 2}\,
(\sqrt{N}\;\xi_N(x)-\sqrt{N+\al}\;\xi_{N-1}(x))\;
(\sqrt{N+\al}\;\ep\xi_N(y)-\sqrt{N}\;\ep\xi_{N-1}(y)).\label{Lag1}\eq

These were established for $\al>0$. For $-1<\al\leq 0$ we must find the
analytic
continuations of the factors in (\ref{Lag4}) and (\ref{Lag1}). The first
factors cause no difficulty since they are defined and analytic for
$\al>-1$.
The same is true of the second factor in (\ref{Lag4}) since \linebreak
$\sqrt{N}\,p_N-
\sqrt{N+\al}\,p_{N-1}$ has zero constant term for $\al>0$ (since
$\psi_1\in\cH$) and
so for all $\al$.

The second factor in (\ref{Lag1}) requires analytic continuation.
Assuming at first that $\al>0$ we write it as
\[-\int_y^{\iy}(\sqrt{N+\al}\;\xi_N(z)-\sqrt{N}\;\xi_{N-1}(z)\,dz
+\hf\int_0^{\iy}(\sqrt{N+\al}\;\xi_N(z)-\sqrt{N}\;\xi_{N-1}(z))\,dz.\]
Now the fact $\ep\ps_1\in\cH$ established earlier implies
that $\int_0^{\iy}\ps_1(z)\,dz=0$. This is equivalent to
\[\sqrt{N}\int_0^{\iy}\xi_N(z)\,dz=\sqrt{N+\al}\int_0^{\iy}\xi_{N-1}(z)\,dz,\]
so the last integral, with its factor $1/2$, is equal to
\[\hf\int_0^{\iy}\Big(\sqrt{N+\al}-{N\ov\sqrt{N+\al}}\Big)\,\xi_N(z)\,dz\]
\[={\al\ov2\sqrt{N+\al}}\int_0^{\iy}z^{\al/2-1}\,e^{-z/2}\,p_N(z)\,dz
=-{1\ov\sqrt{N+\al}}\int_0^{\iy}z^{\al/2}\,(e^{-z/2}\,p_N(z)'\,dz.\]
Hence the second factor in (\ref{Lag1}) is equal to
\bq -\int_y^{\iy}(\sqrt{N+\al}\;\xi_N(z)-\sqrt{N}\;\xi_{N-1}(z))\,dz
-{1\ov\sqrt{N+\al}}\int_0^{\iy}z^{\al/2}\,(e^{-z/2}\,p_N(z))'\,dz.\label{S1factor}\eq
This is analytic for all $\al>-1$ and so provides the desired analytic
continuation.

When $\al=0$ \[L_N(x)-L_{N-1}(x)={x\ov N}L_N'(x),\]
and we find that
\[S^{(4)}_N(x,y)=K_N(x,y)+
{1\ov 2}e^{-x/2}\,L_N'(x)\;\int_0^ye^{-z/2}\,L_N'(z)\,dz,\]
\[S^{(1)}_N(x,y)=K_N(x,y)+
{1\ov 2}e^{-x/2}\,L_N'(x)\;\Big(\int_0^ye^{-z/2}\,L_N'(z)\,dz+1\Big).\]
For both we used the fact
$\int_0^{\iy}\ps_1(z)\,dz=0$ once again and, for the latter, (\ref{S1factor}).\sp

\begin{center}{\bf Acknowledgements}\end{center}

The author thanks Shinsuke M. Nishigaki, who caught an error in the original
version of this paper. Research was supported by the National
Science Foundation through grants DMS-9424292 and DMS-9732687. \sp

\end{document}